\documentstyle[prl,multicol,aps,epsfig]{revtex}
\begin{document}
\title{Single qubit measurements with an asymmetric single-electron
transistor}
\author{S.A. Gurvitz$^{1,2}$ and G.P. Berman$^2$\\
$^1$Department of Particle Physics, Weizmann Institute of Science,
Rehovot 76100, Israel\\
$^2$Theoretical Division, Los Alamos National Laboratory, Los Alamos, NM 87545, USA\\
}
\maketitle
\begin{abstract}
We investigate qubit measurements using a single electron transistor
(SET). Applying the Schr\"odinger equation to the entire
system we find that an asymmetric SET is considerably
more efficient than a symmetric SET. Yet, its efficiency
does not reach that of an ideal detector even in the large asymmetry limit. 
We also compare the SET detector with a point-contact
detector. This comparison allows us to illuminate the relation 
between information gain in the measurement process and 
the decoherence generated by these measurement devices.
\end{abstract}
\hspace{1.5 cm}
\begin{multicols}{2}
An obvious candidate for read-out of the two-state system
(qubit) is the single electron transistor (SET)\cite{dev,schon}. In many
respects it is better than the point-contact (PC) detector\cite{dev}, which
has already been used for quantum measurements\cite{Buks}. 
It has been shown, however, that the symmetric SET
has a rather low sensitivity in its normal working regime\cite{kor1,moz1}. 
Therefore it becomes very important to investigate how to improve the
effectiveness of the SET by a proper selection of its parameters.

In this Letter we examine qubit measurements using the SET
by applying the Schr\"odinger equation to the entire system of the
qubit and detector. In this case, we can unambiguously determine
the backaction of the charge fluctuations in the SET on the qubit
and the sensitivity of the measurement as a function of the
detector parameters. We find that by varying the tunneling
barriers of the SET, one can force the latter to operate in a
regime where the ``active measurement'' time is very short.
Then the SET behaves as a linear quantum detector
even if it is strongly coupled to the measured system. We also demonstrate  
that in this regime the effectiveness of the SET considerably
increases, although it does not reach the ultimate value  
corresponding to an ideal detector. 
By varying the set-up parameters of the SET one can 
investigate decoherence generated by this detector in the operation
mode in which the signal decreases 
to zero. This illuminates the relationship between decoherence and
distinguishability in different types of measurements.

Consider an electrostatic qubit, represented by an electron in a
coupled dot. The qubit is placed in close proximity to the
SET, Fig.~1. The latter is shown as a potential well, coupled to
two reservoirs at different chemical potentials, $\mu_{L,R}$. 
We assume that the bias voltage $V=\mu_L-\mu_R$ is smaller than the 
energy spacing of the single particle states of the well. This condition 
is usually realized in semiconductor quantum dots\cite{averin}.
Then transport takes place through an individual single particle
level, $E_0$, in contrast to a metallic SET, in which many conducting levels
contribute to transport\cite{fn2}. 
One finds that if the electron occupies the upper dot (Figs.~1a,1b)
the current flows through the level $E_0$ of the SET. However, if the
electron occupies the lower dot (Figs.~1c,1d), the SET is blocked
due to the electron-electron repulsion $U$. Here we assume that
$U$ is large enough so that
$E_0+U-\mu_L\gg\Gamma_{L,R}$, where $\Gamma_{L,R}$ are the partial
widths of the level $E_0$ due to coupling with the reservoirs. Thus 
the probability of finding the qubit in the state $E_1$ (Fig.~1a,1b) 
would be given by the ratio, $I(t)/I_0$, 
where $I(t)$ is the (ensemble) average detector current and
$I_0=e\Gamma_L\Gamma_R/(\Gamma_L+\Gamma_R)$ is the detector current
in the absence of a qubit. 
%%%%%%%%%%%%%%%%%%%%%%%%%%%%%%%%%%%%%%%%%%%%%%%%%%%
\begin{figure}
{\centering{\epsfig{figure=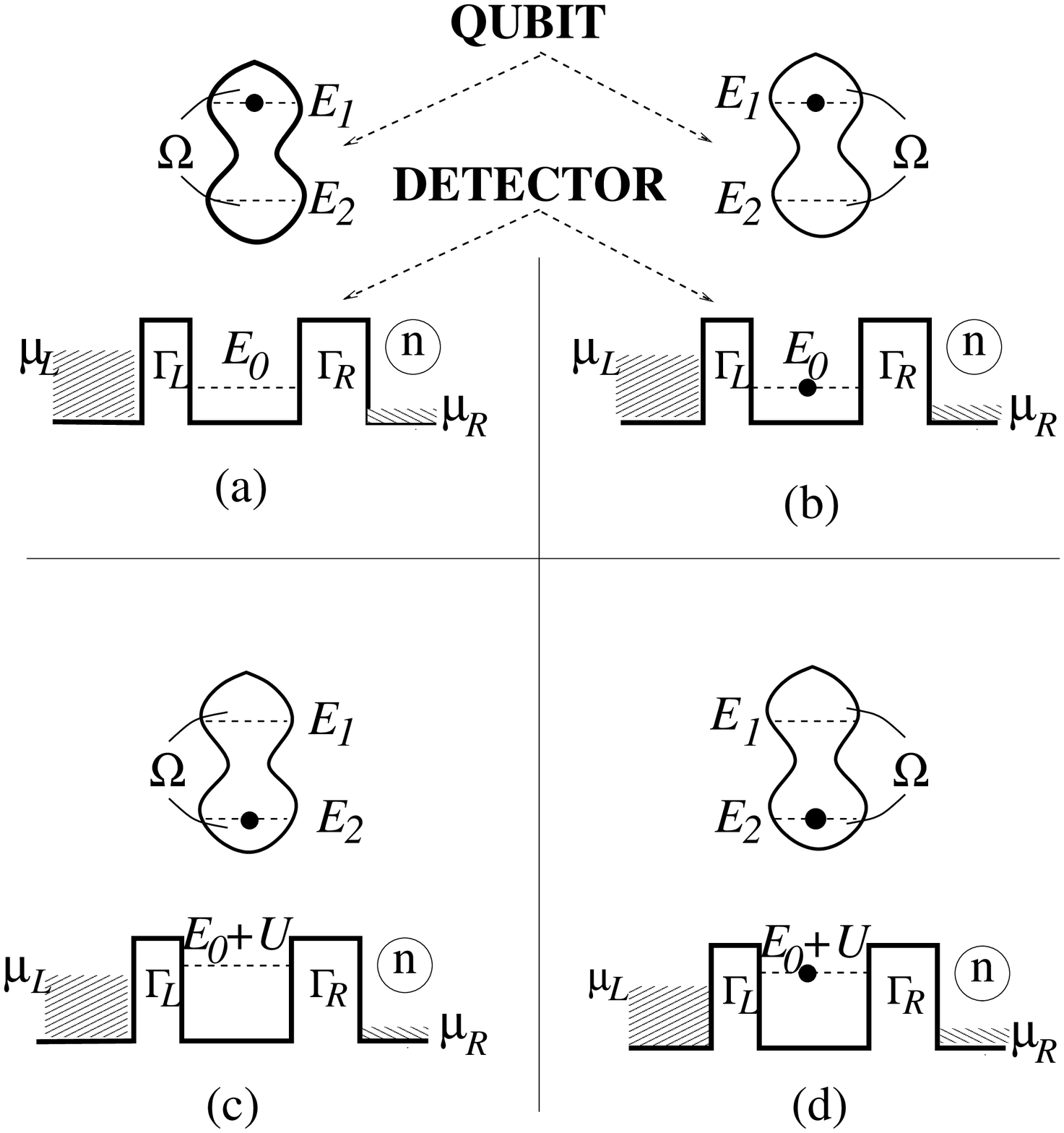,height=9cm,width=8cm,angle=0}}}
{\bf Fig.~1:}
Possible electron configurations of the measured qubit
and the detector. Here $n$ denotes the number of electrons
which have arrived at the right reservoir by time $t$.
\end{figure}
%%%%%%%%%%%%%%%%%%%%%%%%%%%%%%%%%%%%%%%%%%%%%%%%%%%%

The ideal linear measurement, however, cannot be
fully realized since the electric current would flow through the
SET even if the electron of the qubit occupies the lower dot
(Fig.~1d)\cite{fn1}. In addition the state (1b) prevents
qubit oscillations due to the electron-electron
repulsion $U$. As a result the average detector current could
not follow the qubit behavior.

Thus, in order to increase the effectiveness of the
detector, we must diminish the role of the states 1b and 1d
in the measurement dynamics. This can be realized by using 
an asymmetric SET with  $\Gamma_R\gg\Gamma_L$, since the  
probability of finding an electron
inside the quantum well of the SET is $\Gamma_L/(\Gamma_L+\Gamma_R)$.
Then the SET detector spends most of the time in the states 1a and 1c,
where no interaction with the qubit takes the place. This is similar to the
``interaction-free'' measurement, discussed in the
literature\cite{dicke}, which distorts
the measured quantum system in a minimal way.   
Yet, by enhancing the asymmetry of the detector one slows down
the measurement rate of the detector, as well as the fluctuations of
the detector current. Therefore in order to substantiate our qualitative
arguments, one needs to evaluate explicitly the detector efficiency,
which involves the ratio of these two quantities.

Let as consider the entire system as described by the following
tunneling Hamiltonian: 
$H=H_{q}+H_{set}+H_{int}$, where
\begin{eqnarray}
&&H_{q}=E_1a_1^\dagger a_1+E_2a_2^\dagger a_2+\Omega(a_1^\dagger a_2
+a_2^\dagger a_1)~~~~~~~{\mbox{and}}\nonumber\\
&&H_{set}=H_0+E_0c_0^\dagger c_0+\sum_\lambda(\Omega_\lambda^Lc^\dagger_0c_\lambda^L
+\Omega_\lambda^Rc^\dagger_0 c_\lambda^R+H.c.)
\label{a1}
\end{eqnarray}
are the qubit and the SET Hamiltonians, respectively. 
$H_0=\sum_\lambda [E_\lambda^L(c_\lambda^L)^\dagger c_\lambda^L
+E_\lambda^R (c_\lambda^R)^\dagger c_\lambda^R]$ describes
the reservoirs. Here $a^\dagger(a)$ is the creation (annihilation)
operator for the electron in the qubit and $c^\dagger(c)$
is the same operator for the SET; $\Omega$ is the
coupling between the states $E_{1,2}$ of the qubit, and
$\Omega_\lambda^{L,R}$ are the couplings between the level $E_0$ and the
level $E_\lambda^{L,R}$ in the left (right) reservoir. The corresponding
tunneling rates $\Gamma_{L,R}$ are therefore
$\Gamma_{L,R}=2\pi\rho_{L,R}\Omega_{L,R}^2$, where $\rho_{L,R}$
are the density of states in the reservoirs. In addition, we
assumed a weak energy dependence of the couplings,
$\Omega_\lambda^{L,R}\simeq\Omega_{L,R}$. The interaction between the
qubit and the SET is described by $H_{int}=Ua_2^\dagger
a_2c_0^\dagger c_0$

It was demonstrated in Ref.\cite{g} that in the limit of large
$V$ and $U$ (Fig.~1), the Schr\"odinger equation for the
entire system, $i\partial_t|\Psi (t)\rangle =H|\Psi (t)\rangle$,
can be reduced to a Bloch-type rate equation describing the
reduced density-matrix of the entire system, $\sigma_{jj'}^{n}(t)$,
where $j,j'=\{a,b,c,d\}$, and the index $n$ denotes the number of
electrons which have arrived at the right reservoir by the time $t$ (Fig.~1).
This reduction of the Schr\"odinger equation
takes place after partial tracing over the
reservoir states, and it becomes exact in the large bias
limit without any explicit use of any Markov-type or weak coupling
approximations. The diagonal terms of this density matrix
$\sigma_{jj}^{n}(t)$ are the probabilities of finding the
system in one of the states shown in Fig.~1. The off-diagonal
matrix elements (``coherencies''), describe the linear
superposition of these states. As a result we arrive at the following Bloch-type
equations describing the entire system \cite{g,g1}
\begin{mathletters}
\label{a2}
\begin{eqnarray}
\dot\sigma_{aa}^{n}&=&-\Gamma_L\sigma_{aa}^{n}+\Gamma_R\sigma_{bb}^{n-1}
+i\Omega (\sigma_{ac}^{n}-\sigma_{ca}^{n}),\\
\label{a2a}
\dot\sigma_{bb}^{n}&=&-\Gamma_R\sigma_{bb}^{n}+\Gamma_L\sigma_{aa}^{n}
+i\Omega (\sigma_{bd}^{n}-\sigma_{db}^{n}),\\
\label{a2b}
\dot\sigma_{cc}^{n}&=&\Gamma_L\sigma_{dd}^{n}+\Gamma_R\sigma_{dd}^{n-1}
+i\Omega (\sigma_{ca}^{n}-\sigma_{ac}^{n}),\\
\label{a2c}
\dot\sigma_{dd}^{n}&=&-(\Gamma_L+\Gamma_R)\sigma_{dd}^{n}
+i\Omega (\sigma_{db}^{n}-\sigma_{bd}^{n}),\\
\label{a2d}
\dot\sigma_{ac}^{n}&=&i\epsilon\sigma_{ac}^{n}
+i\Omega (\sigma_{aa}^{n}-\sigma_{cc}^{n})
-{\Gamma_L\over 2}\sigma_{ac}^{n}+\Gamma_R\sigma_{bd}^{n-1},\\
\label{a2e} \dot\sigma_{bd}^{n}&=&i\epsilon_1\sigma_{bd}^{n}+
i\Omega (\sigma_{bb}^{n}-\sigma_{dd}^{n}) -\left
(\Gamma_R+{\Gamma_L\over 2}\right )\sigma_{bd}^{n}, \label{a2f}
\end{eqnarray}
\end{mathletters}
where $\epsilon =E_2-E_1$ and $\epsilon_1 =E_2-E_1+U$.

Solving Eqs.~(\ref{a2}) we can determine the time evolution of the
qubit and the detector during the measurement process.
Indeed, the qubit behavior is described by the corresponding
(reduced) density matrix $\sigma_{QB}(t)\equiv 
\{\sigma_{\alpha\alpha'}(t)\}$ with $\alpha,\alpha'=\{1,2\}$,
obtained by tracing the detector variables ($n$):
$\sigma_{11}=\sum_n (\sigma_{aa}^n+\sigma_{bb}^n)$,
$\sigma_{12}=\sum_n
(\sigma_{ac}^n+\sigma_{bd}^n)$.
Here $\sigma_{11}$ and  $\sigma_{22}=1-\sigma_{11}$ are the probabilities
of finding the qubit in the states $E_{1,2}$ respectively, and
$\sigma_{12}$ describes the linear superposition of these states.

The detector behavior, on the other hand, is obtained from
$\sigma_{jj'}^n(t)$ by tracing over the qubit variables. For
instance, the probability of finding $n$ electrons in the
collector at time $t$ is $P_n(t)=\sum_j\sigma_{jj}^{n}(t)$. This
quantity allows us to determine the average detector current,
\begin{equation}
I(t)=e\sum_n n\dot P_n(t)=e\Gamma_R\left [\sigma_{bb}(t)
+\sigma_{dd}(t)]\right ], 
\label{a4}
\end{equation}
where $\sigma_{jj'}=\sum_n\sigma_{jj'}^n$, 
and the shot-noise power spectrum, $S(\omega )$, is given by  
the McDonald formula\cite{moz2},
\begin{equation}
S(\omega) = 2e^2\omega \int_0^\infty dt
\sin (\omega t) \sum_n n^2\dot P_n(t)\, ,
\label{a6}
\end{equation}
One finds from Eqs.~(\ref{a2}), (\ref{a6}) that 
\begin{equation}
S(\omega) = 2e^2\omega\Gamma_R {\mbox{Im}}\,[Z_{bb}(\omega )
+Z_{dd}(\omega )]\, ,
\label{a66}
\end{equation}
where $Z_{jj'}(\omega) = \int_0^\infty \sum_n(2n+1)\sigma_{jj'}^n(t)\exp
(i\omega t)dt$.
%\begin{equation}
%Z_{jj'}(\omega) = \int_0^\infty \sum_n(2n+1)\sigma_{jj'}^n(t)\exp
%(i\omega t)dt\, .
%\label{a67}
%\end{equation}
These quantities are obtained directly from Eqs.~(\ref{a2})
which are reduced to a system of linear algebraic equations
after the corresponding integration over $t$.
%%%%%%%%%%%%%%%%%%%%%%%%%%%%%%%%%%%%%%%%%%%%%%%%%%%
\begin{figure}
{\centering{\epsfig{figure=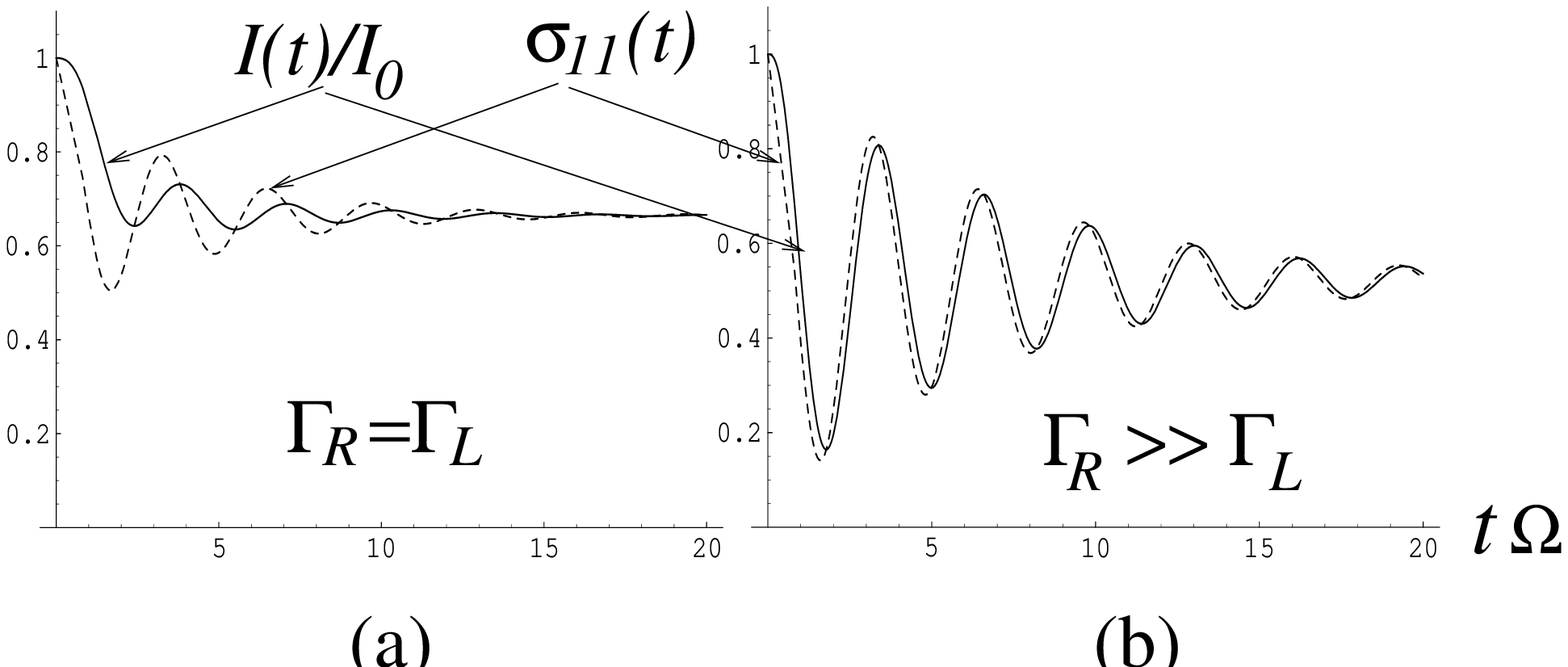,height=3.5cm,width=8.5cm,angle=0}}}
{\bf Fig.~2:}
The probability of finding the qubit electron in the upper dot,
$\sigma_{11}(t)$ (dashed line) in comparison with
$I(t)/I_0$ (solid line) as a
function of the time $t$ for $U=20 \Omega$ and (a) 
$\Gamma_L=\Gamma_R=0.5\Omega$ and (b) 
$\Gamma_L=0.5\Omega$, $\Gamma_R=5\Omega$.
\end{figure}
%%%%%%%%%%%%%%%%%%%%%%%%%%%%%%%%%%%%%%%%%%%%%%%%%%%%

Consider first the average detector current, $I(t)$, Eq.~(\ref{a4}).
We can investigate
the linearity of the measurement\cite{aver1} by comparing the ratio $I(t)/I_0$
with $\sigma_{11}(t)$. These quantities are shown in Fig.~2
for a symmetric SET (a) and an asymmetric SET (b) as a function of time
for $\epsilon =0$.
The initial condition corresponds to the qubit electron in the
upper dot and the detector current $I=I_0$. This implies that
$\sigma_{aa}(0)=\Gamma_R/(\Gamma_L+\Gamma_R)$,
$\sigma_{bb}(0)=\Gamma_L/(\Gamma_L+\Gamma_R)$, while all other 
$\sigma_{jj'}(0)=0$.

One finds from Fig.~2a that for a symmetric SET the 
detector current does not follow the qubit oscillations and 
the qubit behavior is strongly distorted. As we suggested above 
this is caused by the configurations (b) and (d) in Fig.~1.
The result of our calculations for an asymmetric SET, shown in Fig.~2b,
confirms these qualitative arguments. Indeed the weight of these 
configurations is strongly reduced in this case. (In fact,
in the case of $U\gg\Omega,\Gamma$, considered in Fig.~2,
the contribution (d) remains very small even for the symmetric SET). 
As a result, the Rabi oscillations of the qubit are very well
reproduced by the ratio $I(t)/I_0$ and the qubit is distorted much
less than in the case of the symmetric SET.
%%%%%%%%%%%%%%%%%%%%%%%%%%%%%%%%%%%%%%%%%%%%%%%%%%%
\begin{figure}
{\centering{\epsfig{figure=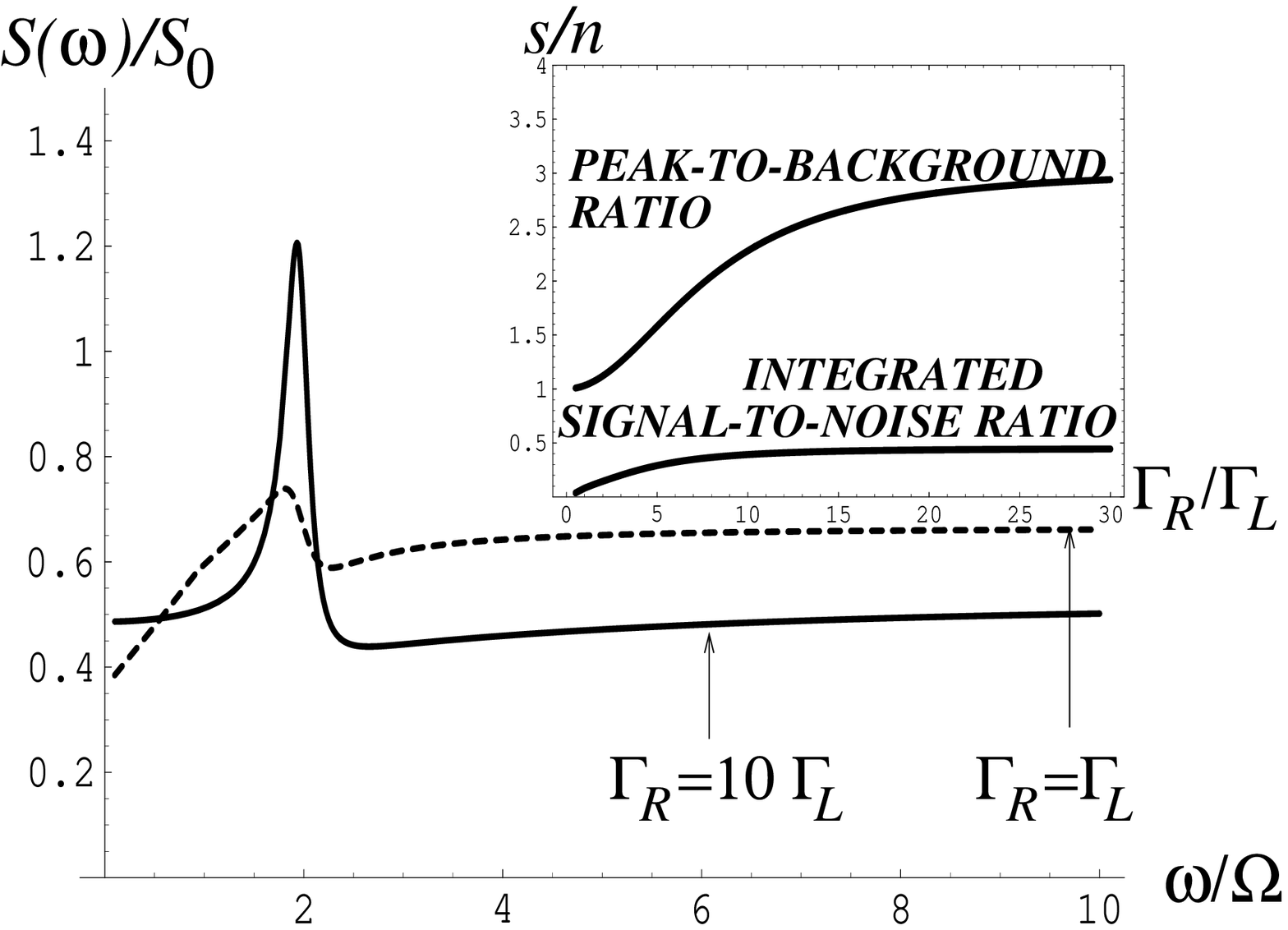,height=6cm,width=8.5cm,angle=0}}}
{\bf Fig.~3:}
Shot-noise power spectrum for the symmetric (dashed line)
and the asymmetric (solid line) SET. The parameters are the same
as in Fig.~2. The peak to-background ratio 
and the integrated signal-to-noise ratio are
shown in the inset as a function of $\Gamma_R/\Gamma_L$.
\end{figure}
%%%%%%%%%%%%%%%%%%%%%%%%%%%%%%%%%%%%%%%%%%%%%%%%%%%%

The result shown in Fig.~2 demonstrates the linearity of measurement
achieved by the asymmetric SET. Yet, the ratio $I(t)/I_0$ does not
yield the effectiveness of measurement. The latter is reflected 
in the detector noise spectrum (Fano factor), $S(\omega )/S_0$,
where $S(\omega )$ is given by Eqs.~(\ref{a6}), (\ref{a66}), and
$S_0=2eI_0$. The results for symmetric and asymmetric SET are shown
in Fig.~3. One finds that the peak in the spectrum corresponding
to the Rabi frequency of the qubit is much more pronounced for
the asymmetric SET. It was argued in\cite{kor1,goan} that a
measure of a measurement efficiency is the peak-to-background ratio
of the detector current power spectrum, with a perfectly efficient detector
yielding a ratio of 4. This quantity is shown in the inset of Fig.~3 as a function 
of the detector asymmetry. One finds that it approaches the value of 3. 
Thus an increase of the asymmetry considerably improves the efficiency
of the SET although it does not reach the effectiveness of an ideal detector.

In addition, the effectiveness of measurement can be represented by
the integrated signal-to-noise ratio\cite{brag,moz1}
$s/n=\int_{-\infty}^\infty [|I_{sig}(\omega )|^2/
S(\omega )] d\omega / 2\pi$. 
In our case the signal corresponds to the deviation of the detector
current from its stationary value. Thus,
$I_{sig}(\omega )=\int_0^\infty \Delta I(t)\exp (i\omega t)dt$,
where $\Delta I(t)=I(t)-I(t\to\infty )$. 
Using Eqs.~(\ref{a2}), (\ref{a4}) and  (\ref{a66}) one can evaluate
the integrated signal-to-noise ratio as a function of
$\Gamma_R/\Gamma_L$. The results of our calculations for $\epsilon =0$,
$\Gamma_L=0.5\Omega$ and $U=20\Omega$ are shown in the inset
of Fig.~3. As expected, the signal-to-noise ratio increases with
$\Gamma_R/\Gamma_L$.

Although our calculations dealt with a
strongly coupled detector, we anticipate that the asymmetric SET
would be  more effective than the symmetric one 
even for a weak coupling or for a metallic SET.
Indeed the weight of the configurations
(b) and (d) in Fig.~1 is always suppressed for $\Gamma_R\gg\Gamma_L$.
As a result, the asymmetric SET would distort
the measured system less than the symmetric one. In fact,
this seems to be at variance with Refs.\cite{averin},
which predict high effectiveness of the {\em symmetric} SET
in the weak coupling regime. Yet recent
analysis\cite{moz1} that demonstrates low effectiveness
of a symmetric SET in the weak coupling regime, supports our
qualitative arguments. In addition, Refs.\cite{kor1} also predict
low effectiveness of the symmetric SET in the weakly responding regime.   

In is interesting to compare qubit measurements using SET 
and PC detectors. The latter, considered as an ideal
detector can be schematically represented
by a tunneling barrier separating the left and the right reservoirs
(instead of the two tunneling barriers in the SET, Fig.~1). The height
of this barrier is modulated by the electrostatic field of the
electron in the qubit. As a result, the current flowing through
the point contact varies from $I_1$, when the electron occupies the
level $E_1$ of the qubit, to $I_2$, if the electron occupies the
level $E_2$ (Fig.~1). One finds that in the case of $E_1=E_2$ 
the qubit density matrix, $\sigma_{QB}(t)$
is described by the following Bloch-type rate equations 
\begin{mathletters}
\label{c1}
\begin{eqnarray}
\dot\sigma_{11}&=&i\Omega (\sigma_{12}-\sigma_{21}),\\
\label{c1a}
\dot\sigma_{12}&=&i\Omega(2\sigma_{11}-1)
-(\Gamma_d/2)\sigma_{12},
\label{c1b}
\end{eqnarray}
\end{mathletters}   
with the decoherence rate
$\Gamma_d=(\sqrt{I_1/e}-\sqrt{I_2/e})^2$\cite{gur}.
This result supports the idea that, for the case of an ideal detector,  
the measurement-induced decoherence rate  
is directly related to the detector signal, or more
precisely, to the ability of the detector to distinguish between
different states of the qubit\cite{aver1,clerk}. 
Indeed, the same expression for the decoherence rate
can be obtained using a Bayesian formalism that shows the
tendency of the qubit to evolve corresponding to the information
acquired from an ideal measurement\cite{kor1}.

It is interesting that in the case of asymmetric SET detector
and large $U$ the qubit density matrix is described by the same
Eqs.~(\ref{c1}) as in the case of PC (ideal) detector. 
This can be obtained by tracing over $n$ in Eqs.~(\ref{a2})
and neglecting small terms of the order of $\Gamma_L/\Gamma_R$,
$\Gamma_{L,R} /U$ and $\Omega /U$. As a result, one arrives to Eqs.~(\ref{c1}),
with the decoherence rate $\Gamma_d=\Gamma_L$. The latter
corresponds to $I_1=I_0\to e\Gamma_L$ and $I_2=0$. Thus,  
$\Gamma_d$ is directly related to the detector signal as
in the case of the PC detector\cite{kor1}. Nevertheless
the asymmetric SET does not become an ideal one, even in the large asymmetry limit,
$\Gamma_L/\Gamma_R\ll 1$. Indeed, the
peak-to-background ratio reaches the value 3 for the asymmetric SET (Fig.~3),
whereas it becomes 4 for the PC detector\cite{kor1,goan,fn3}.

The above result might question the possible connection of decoherence with
information gain. Indeed, it is assumed that in the case of a
non-ideal detector (the SET) a part of the decoherence rate
is generated by ``pure environment''\cite{kor1}, in addition to that related to the
measurement rate\cite{aver1,clerk}. Yet, it follows from our analysis
that the decoherence rate generated by the asymmetric SET
is given by the same expression as that of the PC detector, i.e. it is related 
to the measurement only. Next, one could anticipate that
the decoherence rate diminishes with the detector signal, since
then only the ``pure environment'' would contribute to decoherence. 
This can be expected, for instance by increasing     
the chemical potential $\mu_L$ in Fig.~1, so that $\mu_L\gg E_0+U$. As a result
the detector current becomes the same for the both states of the qubit: 
$I_1=I_2=I_0\simeq e\Gamma_L$. This case can be studied with
the corresponding rate equations, similar to Eqs.~(\ref{a2}).
As in the previous case one finds that in the limit of
$\Gamma_L\ll\Gamma_R$ and $U\gg\Gamma_{L,R}$ the qubit 
density matrix is given by Eqs.~(\ref{c1}), but with $\Gamma_d=2\Gamma_L$. 

Such an increase of $\Gamma_d$ in the no-measurement regime can 
be understood if the decoherence generated by the SET is associated 
with the detector shot-noise, rather than the information flow. 
Indeed, the shot-noise increases by a factor of two with respect to the
measurement case. The reason is that in the measurement regime the
detector is blocked whenever the electron occupies the
state $E_2$ of the qubit, Fig.~1. In any case, irrespective of any possible 
interpretation, our results demonstrate that the reduction 
of information acquired by a measurement,
may not lead to a decrease of the decoherence rate, even for 
a nearly ideal detector.
 
In summary, we have found that the effectiveness of the SET
detector can be considerably increased by making it asymmetric,
where the tunneling to the collector is much faster
than the tunneling to the emitter reservoir. In this
case the effectiveness of the detector becomes more like that
of the PC detector. Yet the decoherence rate 
behaves differently when no signal can be extracted from the measurement. 
In particular, the decoherence rate increases by a factor of two in the 
case of the SET, whereas it vanishes for the PC detector.
This questions the role of information gain
in the decoherence mechanism. We believe that
this result should be taken into account in any consideration of
how to diminish the decoherence of a single quantum system interacting
with a detector.

One of the authors (S.G.) acknowledges useful discussions with D.
Mozyrsky. This work was supported by the Department of Energy
(DOE) under Contract No. W-7405-ENG-36, by the National Security
Agency (NSA), and by the Advanced Research and Development
Activity (ARDA) contract under Army Research Office  (ARO) \# 707003.

\end{multicols}

\begin{references}
\bibitem{fn2} In fact, our treatment remains essentially the same for many 
conducting levels contributing to transport, providing that their widths do
not overlap (c.f. with\cite{schon}). Otherwise, the
treatment should be modified.
\bibitem{fn1} Although the state 1d is not directly accessible due to  
transitions from the left reservoir, it can reached in the course of the Rabi 
oscillations of the qubit if $\Omega$ and $U$ are comparable.
\bibitem{fn3} It was argued, however, in Refs.\cite{kor1} that in the
case of a weak response the asymmetric SET becomes an ideal detector
for $\Gamma_R/\Gamma_L\gg 1$.
\end{references}
\end{document}